# Machine Learning Interatomic Potential for Anisotropic Thermal Transport in Bulk Hexagonal Boron Nitride

Jialin Tang[1,2], Qi Wang[1], Jiongzhi Zheng[3], Lin Cheng[1,*], Ruiqiang Guo[1,*]

[1]*Thermal Science Research Center, Shandong Institute of Advanced Technology, Jinan, Shandong 250103, China*
[2]*Institute of Advanced Technology, Shandong University, Jinan, Shandong 250061, China*
[3]*Department of Mechanical and Aerospace Engineering, The Hong Kong University of Science and Technology, Clear Water Bay, Kowloon, Hong Kong*

**ABSTRACT**

The highly anisotropic thermal conductivity in layered materials is crucial for a broad range of applications such as thermal management of electronic devices, thermal insulation, and thermoelectrics. Understanding of anisotropic thermal transport in layered materials largely depends on atomistic simulations based on density functional theory (DFT) or empirical potentials, which however suffer either low computational efficiency or accuracy. Recently, machine learning interatomic potentials (MLIPs) are emerging as a powerful tool to bridge the gap. Despite the recent progress in developing MLIPs, little attention has been paid to constructing a potential that can accurately predict the thermal properties of layered materials, which is more challenging compared with the case of isotropic materials because of the highly anisotropic bonding and weak van der Waals interactions in layered materials. Here, we introduce a MLIP within the Gaussian approximation potential (GAP) framework for bulk hexagonal boron nitride (h-BN) with a typical layered structure. The GAP can well predict the highly anisotropic phonon transport properties and thermal conductivity of bulk h-BN with DFT-level accuracy at orders of magnitude reduced cost. Our work demonstrates the ability of GAP to reproduce the subtle features of anisotropic potential energy surfaces of bulk h-BN and potentially other layered materials. Atomistic simulations based on MLIPs are expected to be able to greatly promote the understanding of phonon transport and the prediction of thermophysical properties in layered materials.

[*] cheng@sdu.edu.cn
[*] rqguo@connect.ust.hk

# I. INTRODUCTION

Thermal transport in layered materials is typically anisotropic due to strong intralayer covalent bonds and weak interlayer van der Waals bonds. The highly anisotropic thermal conductivity makes layered materials very attractive in a variety of applications [1-17]. For example, graphite [4-6] and bulk hexagonal boron nitride (h-BN) [7] have extremely high in-plane thermal conductivity and much lower cross-plane thermal conductivity, making them ideal heat spreaders that can efficiently dissipate heat along the in-plane direction while protecting the devices below them from overheating for thermal management of electronic devices. Because of the high in-plane thermal conductivity, h-BN [8,9] and graphite [10,11] have been used as thermal conducting fillers to enhance the overall thermal conductivity of thermal interface materials. Also, many layered materials such as $Bi_2Te_3$ [12], SnSe [13,14], and BiCuSeO [15] exhibit high thermoelectric performance largely due to the high lattice anharmonicity and thus low thermal conductivity. In addition, the low cross-plane thermal conductivities of BN nanosheets [16] and $MoS_2$ films with random interlayer rotations [17] make them potential thermal insulation materials.

Atomistic simulations have been widely used to understand thermal transport in layered materials. Two routine atomic-scale modeling techniques are *ab initio* calculations based on density functional theory (DFT) [18] and molecular dynamics (MD) [19,20] simulations using empirical potentials (EPs) [21]. For example, DFT combined with the Boltzmann transport equation (BTE) can accurately predict the lattice thermal conductivity and modal phonon properties of crystalline materials [22-

25]. MD simulations based on EPs have been widely used to model thermal processes and calculate thermal properties in various systems [26-28]. Despite the success of conventional simulation techniques, accurately and efficiently describing interatomic interactions remains a great challenge. Although DFT calculations can describe interatomic interactions with high precision, they are computationally expensive, which limits the simulation systems to hundreds of atoms on the picosecond time scales. In contrast, MD simulations using EPs are usually computationally affordable yet less accurate and less transferable than DFT calculations [29,30].

Machine learning is emerging as a powerful tool to construct accurate and efficient interatomic potentials. Extensive efforts have been devoted to the development of machine learning interatomic potentials (MLIPs) in recent years, employing algorithms including Gaussian process regression [31], linear regression [32,33], artificial neural networks [34], kernel ridge regression [35,36], etc. MLIPs have been successfully applied to model thermal properties of several materials, such as amorphous Si [37], crystalline Si with vacancies [38], single-layer $MoS_{2(1-x)}Se_{2x}$ alloys [39], wurtzite BAs [40], $β$-$Ga_2O_3$ [41], LiF and FLiBe molten salts [42], and h-BN monolayer [43]. All these MLIPs achieve a DFT-level accuracy while the computational cost is orders of magnitude lower than that of DFT. To date, MLIPs for thermal transport in layered materials have been rarely reported. Compared with isotropic materials, it is more challenging to develop high-quality MLIPs for layered materials due to the anisotropic topology of the potential energy surfaces (PESs) and the subtle van der Waals interaction that is weak and long-ranged.

In this work, we develop a MLIP based on Gaussian process regression, namely Gaussian approximation potential (GAP) for bulk h-BN, a typical example of layered materials. As a wide gap semiconductor, bulk h-BN has been exploited in various applications, such as solid-state neutron detectors [44,45], single-photon emitters [46,47], supercapacitors [48], ultraviolet light emitters [49], and electron emitters [50]. For most of these applications, the thermal conductivity of h-BN plays a key role in determining the heat dissipation and performance of the system. Also, bulk h-BN is one of the most promising thermal management materials for modern electronics because of its highly anisotropic thermal conductivity and electrical insulation [7-9,16,51]. Its thermal conductivity at room temperature is ~420 and ~4.8 $Wm^{-1}K^{-1}$ along the in-plane and cross-plane directions, respectively [7].

We show that GAP can accurately predict the highly anisotropic thermal conductivity of bulk h-BN, demonstrating the precise description of both strong intralayer covalent bonds and weak interlayer van der Waals bonds. Moreover, the phonon transport properties predicted by the GAP including phonon dispersion, group velocities, volumetric specific heat, and three-phonon scattering rates are all in good agreement with the DFT benchmarks. Particularly, we find that virial stresses play crucial roles in constructing the high-quality GAP model and strongly affect its accuracy in predicting the lattice parameters and force constants of bulk h-BN. This work demonstrates that GAP can serve as a powerful tool to model thermal properties of bulk h-BN, which is expected to be able to promote the understanding of phonon transport and the prediction of thermophysical properties of bulk h-BN and potentially other layered materials.

## II. METHODS

### A. Gaussian Approximation Potential

Gaussian approximation potential method uses a set of descriptors to accurately describe local atomic environments and performs the PESs fitting based on the Gaussian process regression. In this work, the total energy for the GAP sums over the contributions of two-body (2b), three-body (3b), and many-body (MB) interactions [52,53]:

$$E = \delta^{(2b)} \sum_{ij} \varepsilon^{(2b)}(\mathbf{q}_{ij}^{(2b)}) + \delta^{(3b)} \sum_{ijk} \varepsilon^{(3b)}(\mathbf{q}_{ijk}^{(3b)}) + \delta^{(MB)} \sum_{i} \varepsilon^{(MB)}(\mathbf{q}_{i}^{(MB)}) \quad , \quad (1)$$

where $\delta$ represents the weighting factor, $\mathbf{q}$ is a local atomic environment descriptor, the $i$, $j$, and $k$ represent atom indices in the system. The local atomic energy contribution $\varepsilon^{(d)}(\mathbf{q}_i^{(d)})$ is given by a linear combination of kernel functions $K^{(d)}$ and its weight coefficient $\alpha_t^{(d)}$:

$$\varepsilon^{(d)}(\mathbf{q}_i^{(d)}) = \sum_{t=1}^{N_t^{(d)}} \alpha_t^{(d)} K^{(d)}(\mathbf{q}_i^{(d)}, \mathbf{q}_t^{(d)}) \quad , \quad (2)$$

where $d$ is the corresponding descriptor, the kernel function $K^{(d)}(\mathbf{q}_i^{(d)}, \mathbf{q}_t^{(d)})$ is a measure of the similarity between the previously observed atomic environment $\mathbf{q}_t^{(d)}$ and the to-be-predicted atomic environment $\mathbf{q}_i^{(d)}$.

In this work, we use the Smooth Overlap of Atomic Positions (SOAP) descriptor to describe many-body interactions. The SOAP descriptor represents the atomic environment using local atomic neighbor density $\rho_i(\mathbf{r})$ resulting from the summation of the Gaussian function placed on each neighbor atom within a cutoff radius ($r_{cut}$).

The density $\rho_i(\mathbf{r})$ can be expanded in a basis set of the radial function $g_{(n)}(r)$ and the spherical harmonic function $Y_{lm}(\mathbf{r})$:

$$\rho_i(\mathbf{r}) = \sum_{nlm} c_{nlm}^{(i)} g_{(n)}(r) Y_{lm}(\mathbf{r}). \qquad (3)$$

Here, $c_{nlm}^{(i)}$ are the expansion coefficients for atom *i*.

**B. Construction of the Training and Testing Databases**

We constructed the training and testing databases using the total energy, atomic forces, and/or virial stresses of bulk h-BN configurations selected from the *ab initio* calculations. For both the training and testing databases, each configuration contains 200 atoms in a 5 × 5 × 2 supercell. Firstly, we ran *ab initio* molecular dynamics (AIMD) simulations in the canonical ensemble (*NVT*: constant number of atoms, volume, and temperature) at 300 K with a time step of 0.5 fs via the Vienna Ab-Initio Simulation Package (VASP) [54]. Then, the configurations for the training and testing databases were generated by selecting one snapshot every 75 time steps. Next, the self-consistent field (SCF) calculations implemented in the VASP package were performed on the selected configurations to record the total energies, atomic forces, and virial stresses. Both AIMD simulations and SCF calculations were performed within the local density approximation (LDA) [55] with the projector augmented wave method [56] and the cutoff energy was set to 600 eV. LDA has been demonstrated to be able to accurately predict the lattice thermal conductivity of layered materials, such as $Bi_2Te_3$ [57], $Bi_2Se_3$ [58], SnSe [13], and $HfTe_5$ [59]. We employed 1 × 1 × 1 and 2 × 2 × 2 k-point grids to sample the Brillouin zone for AIMD and SCF calculations, respectively. The energy convergence criteria for both AIMD and SCF calculations are set to be $10^{-8}$ eV/atom and the force convergence criteria for SCF calculations are set to be $10^{-7}$ eV/Å.

**C. Thermal Transport Calculations**

To obtain the thermal conductivity of bulk h-BN, we first calculated the second-order and third-order interatomic force constants (IFCs) from the GAP and DFT using the finite displacement method with a 5 × 5 × 2 supercell. For the calculations of the third-order IFCs, we considered up to the 12th nearest neighbors to ensure the convergence of the thermal conductivity. The lattice thermal conductivities of bulk h-BN were calculated by iteratively solving linearized BTE using the ShengBTE software [24], with a 24 × 24 × 9 q-mesh for sampling the Brillouin zone. Within the framework of the BTE, the thermal conductivity can be expressed as,

$$k_{\alpha\beta} = \frac{1}{\Omega} \sum_{\lambda} C_V(\lambda) v_g^{\alpha}(\lambda) v_g^{\beta}(\lambda) \tau(\lambda), \qquad (4)$$

where $\alpha$ and $\beta$ are the components of the thermal conductivity tensor, $\Omega$ is the volume of the unit-cell. $v_g(\lambda)$, $C_V(\lambda)$, and $\tau(\lambda)$ are the group velocity, the volumetric specific heat, and the lifetime of each phonon mode $\lambda$, respectively.

**D. Interaction Strength Described by the Second-order IFCs**

The cutoff radius for GAP training can be estimated by the interatomic interaction strength, which is described by the second-order IFCs in this work. Using the finite displacement method, the second-order IFCs tensor is computed by

$$\Phi_{ij}^{\alpha\beta} = \frac{\partial^2 E}{\partial u_i^\alpha \partial u_j^\beta} = \begin{bmatrix} \frac{\partial^2 E}{\partial u_i^x \partial u_j^x} & \frac{\partial^2 E}{\partial u_i^x \partial u_j^y} & \frac{\partial^2 E}{\partial u_i^x \partial u_j^z} \\ \frac{\partial^2 E}{\partial u_i^y \partial u_j^x} & \frac{\partial^2 E}{\partial u_i^y \partial u_j^y} & \frac{\partial^2 E}{\partial u_i^y \partial u_j^z} \\ \frac{\partial^2 E}{\partial u_i^z \partial u_j^x} & \frac{\partial^2 E}{\partial u_i^z \partial u_j^y} & \frac{\partial^2 E}{\partial u_i^z \partial u_j^z} \end{bmatrix}, \tag{5}$$

wherein $\alpha$ and $\beta$ represent the directions of atomic displacement of atom *i* and *j*, respectively. The interaction strength is defined as the root mean square (RMS) of all the elements of the second-order IFCs tensor

$$\text{RMS}(\Phi_{ij}) = \left[\frac{1}{9}\sum_{\alpha\beta}(\Phi_{ij}^{\alpha\beta})^2\right]^{\frac{1}{2}}. \tag{6}$$

## III. RESULTS

### A. Training the GAP

Depending on the requirement for the accuracy of MLIPs, the training database may contain total energies, atomic forces, and/or virial stresses. We start by training the GAP using total energies and atomic forces only. To develop an accurate and efficient GAP, we first test the convergence of the GAP model with respect to the cutoff radius ($r_{cut}$), the radial ($n_{max}$) and angular ($l_{max}$) basis set expansion of the atomic neighbor density, and the number of training configurations. Here, we evaluate the accuracy of the GAP model using the root-mean-squared errors (RMSEs) of atomic forces calculated by the GAP with respect to the DFT benchmark for a testing database. The testing database contains 40 configurations that are not included in the training database. The initial training database for the convergence test contains 100 configurations selected from AIMD simulations at 300K.

Figure 1(a) shows the RMSE of atomic forces and relative computational time as a function of the cutoff radius, which is set to be the same for the 2b, 3b, and SOAP descriptors. The relative computational time corresponding to the converged test result is set to be 1. The cutoff radius should be large enough so that necessary interatomic interactions are included in the local atomic environment. However, a too large cutoff radius will cause high computational cost and overfitting for GAP [60]. Considering that the interlayer spacing of bulk h-BN is 3.22 Å, the initial value of the cutoff radius is set to be 3.5 Å so that the atomic interactions between neighboring layers can be included. We find that the RMSEs of atomic forces on B and N atoms reach converged values at a cutoff radius of 4.5 Å, which are 0.01 eV/Å and 0.009 eV/Å, respectively. Choosing 4.5 Å as the cutoff radius is also rationalized by the interaction strength described by the RMS of the second-order IFC tensors RMS($\Phi_{ij}$), which becomes negligible after 4.5 Å. Also, further increasing the cutoff radius results in substantially increased computational cost. Specifically, as the cutoff radius further increases from 4.5 to 5.5 Å, the computational cost almost doubles. Later we will show the GAP constructed using a cutoff radius of 4.5 Å can accurately predict the phonon transport properties and thermal conductivity along the cross-plane direction in bulk h-BN, demonstrating the accurate description of the van der Waals interactions.

In Fig. 1(b), we show the RMSE of atomic forces and relative computational time as a function of the $n_{max}$ and $l_{max}$, which determine the length of the SOAP vector. The $n_{max}$ and $l_{max}$ should be large enough to accurately describe local atomic environments. As $n_{max}$ and $l_{max}$ increase, the RMSE of atomic forces on B and N atoms decreases and reaches converged values at $n_{max}=l_{max}=8$. Further increasing $n_{max}$ and $l_{max}$ results in little improved performance yet notably increased computational cost, e.g., almost doubled as the $n_{max}$ and $l_{max}$ increase from 8 to 12. We therefore used $n_{max}=l_{max}=8$ in all the subsequent GAP training.

Figure. 1(c) shows the RMSE of atomic forces and relative computational time as a function of the number of training configurations. To ensure the high accuracy and transferability of GAP, the representative configurations used for training are expected to be able to cover the entire space of local atomic environments. The RMSEs of atomic forces on B and N atoms are converged at 150 training configurations, which are 0.01 eV/Å and 0.009 eV/Å, respectively. The computational cost increases almost linearly with the increasing number of configurations. Considering the balance between accuracy and computational cost, we finally choose 200 configurations for GAP training.

We summarize the converged test results and other main hyperparameters used for training in Table I. We next use the GAP model trained using these hyperparameters (hereafter named GAP-noStress) for further calculation. As shown in Table II, the lattice constants $a$ and $c$ of bulk h-BN predicted by the GAP-noStress are 2.48 and 6.67 Å, respectively, which are 0.14% smaller and 3.57% larger than the DFT results, respectively. Although the prediction error in lattice constant $a$ is acceptable, that in lattice constant $c$ is too large. In addition, the GAP-noStress largely underestimates the cross-plane thermal conductivity by 55% at 300K relative to the DFT benchmark, indicating the IFCs cannot be precisely predicted.

To improve the accuracy of GAP, we further included the virial stresses in the training data to retrain the GAP model. The expected error of virial stresses ($\sigma_v^{virial}$) is set to be 0.001 eV/Å$^3$. Similarly, we performed the convergence test for the cutoff radius, $n_{max}$, $l_{max}$ and the number of training configurations. As shown in Fig. 2, all three tests show similar trends as those before including the virial stresses in the training data. The RMSEs of atomic forces reach convergence at the same values for all the tested parameters. We therefore used exactly the same hyperparameters to retrain the GAP after including the virial stresses in the training data, herefoth named GAP-Stress. As expected, the corresponding training time increases after including the virial stresses in the training data. For example, the training time of the GAP-stress almost doubles compared to the GAP-noStress case. Figure 3 compares the total energies and atomic forces calculated by the GAP-Stress and DFT of both training and testing databases. For the training database, the RMSEs of total energies and atomic forces on B and N atoms are 0.052 meV/atom, 0.01 eV/Å, and 0.009 eV/Å, respectively. As shown in Fig. 3(b), the testing database has comparable RMSEs of total energies (0.051 meV/atom) and atomic forces on B (0.01 eV/Å) and N atoms (0.009 eV/Å), demonstrating high accuracy of the GAP for the unseen data. We also evaluated the performance of the GAP recently developed using multiple phases of h-BN (hereforth named GAP-Ref) [61] for bulk h-BN, obtaining much larger RMSEs of 0.096 and 0.091 eV/Å for its atomic forces on B and N atoms, respectively.

Although the RMSEs of atomic forces calculated by the GAP-Stress are not notably reduced compared to those calculated by the GAP-noStress, the GAP-Stress results in much smaller prediction errors in lattice constants $a$ (0.04%) and $c$ (0.02%) compared to the GAP-noStress, as shown in Table II. The reduced prediction errors of lattice constants demonstrate the improved accuracy of the GAP-Stress in describing both intralayer and interlayer bonds of bulk h-BN.

## B. Phonon Transport in Bulk h-BN

We now move on to evaluate the accuracy of the GAP-Stress in predicting the thermal properties of bulk h-BN. In Fig. 4, we present the thermal conductivities of bulk h-BN predicted by the GAP as a function of temperature, in comparison with those calculated by the GAP-Ref and DFT. As shown in Fig. 4, the GAP performs well in predicting both the in-plane and cross-plane thermal conductivities within the temperature range from 150 to 1000 K. The in-plane thermal conductivities predicted by the GAP are slightly lower by ~2.9% than the reference DFT data. In contrast, the in-plane thermal conductivities predicted by the GAP-Ref show remarkably different temperature dependence, as indicated by the rapidly increasing deviation from the DFT results with the decrease of temperature. For example, at 150 K, the in-plane thermal conductivity predicted by the GAP-Ref is 14% higher than that predicted by DFT. Along the cross-plane direction, the GAP underpredicts the thermal conductivities by ~7.9% compared to the DFT benchmarks. In contrast, the cross-plane thermal conductivities predicted by the GAP-Ref are ~40% lower than the DFT benchmarks. This large deviation can be attributed to the lower accuracy of the GAP-Ref in describing interatomic interactions, as indicated by the much larger RMSEs of atomic forces presented above.

Figure 5 shows the cumulative thermal conductivity along in-plane and cross-plane directions as a function of phonon mean free path (MFP). The MFP spectrum is a metric of size dependence of thermal conductivity, which, for example, can be used to estimate the effective thermal conductivity of nanomaterials with specific characteristic sizes. We take the results at 150 and 300 K as examples to compare the cumulative thermal conductivity predicted by the GAP, GAP-Ref, and DFT. As shown in Fig. 5, along both the in-plane and cross-plane directions, the cumulative thermal conductivities at 150 and 300 K predicted by the GAP agree well with the DFT benchmarks. In contrast, the GAP-Ref predicts reasonable cumulative thermal conductivities along the in-plane direction but largely deviated spectra along the cross-plane direction at both 150 and 300 K, in comparison with the DFT benchmarks. Specifically, the cross-plane results predicted by the GAP-Ref exhibit a much weaker MFP dependence, which begins at ~130 nm at 150 K and ~45 nm at 300 K, respectively.

We continue to look into phonon transport details that determine thermal conductivity, specifically for the volumetric heat capacity, phonon group velocity, and phonon lifetime. Figure 6(a) shows the volumetric heat capacity calculated by the GAP and GAP-Ref both agree well with the DFT results below 1000 K. Figure 6(b) shows the Grüneisen parameters calculated by the GAP, GAP-Ref, and DFT as a function of temperature. The Grüneisen parameters calculated by the GAP are close to those by DFT, which indicates that the GAP can well predict the third-order IFCs. In contrast, those predicted by the GAP-Ref are much larger than the DFT benchmarks.

Figure 7 further shows the three-phonon scattering rates at 300K, phonon dispersions, and group velocities calculated by the GAP, GAP-Ref, and DFT. The modal three-phonon scattering rates and phonon dispersions calculated by the GAP are in good agreement with the reference DFT data, indicating that the GAP can predict third- and second-order IFCs with high precision. The GAP-Ref can well predict the three-phonon scattering rates of most modes and the phonon dispersion along the in-plane directions, which rationalizes the good prediction of the in-plane thermal conductivity. However, along the cross-plane (ΓA) direction, the six lowest phonon branches predicted by the GAP-Ref are substantially softer than the DFT benchmark [Fig. 7(c)]. For example, the highest frequency of the 6th phonon branch predicted by the GAP-Ref is 2.19 THz at point P (marked by the solid circle), which is 38% lower than that predicted by DFT (3.58 THz). Moreover, the GAP-Ref predicts even opposite frequency dependence for the 6th phonon branch, which first increases from 1.79 THz at $\Gamma$ to 2.19 THz at P, then decreases to 2.15 THz at A, in contrast to the monotonically decreasing frequency (from 3.59 to 2.54 THz) predicted by DFT. The softening of the six phonon branches from GAP-Ref results in substantially lower group velocities, as shown in Figure 7(d). Specifically, the group velocity of the 6th phonon branch from the GAP-Ref first increases from 0 (point $\Gamma$) to 1010 ms$^{-1}$, then decreases to 0 ms$^{-1}$ at P while that predicted by DFT monotonically increases from 0 (point $\Gamma$) to 2363 ms$^{-1}$ (point A). Also, the GAP-Ref predicts increasing group velocity from 0 ms$^{-1}$ at $\Gamma$ to 1131 ms$^{-1}$ at A for the two degenerate TA branches, which contradicts the decreasing one from 1570 to 1123 ms$^{-1}$ calculated by DFT. Considering that the six lowest phonon branches contribute ~86.5% to the overall cross-plane thermal conductivity, the largely underestimated cross-plane thermal conductivity predicted by the GAP-Ref can be attributed to these softened phonon modes along the cross-plane direction.

## C. Transferability of the GAP

MLLPs are expected to have good transferability so that they can be used to model previously unknown scenarios and structures. To evaluate the transferability of our GAP, we first compare the trajectory of potential energy generated by AIMD and classical MD simulations based on the GAP using a 3 ×3 × 2 supercell and a 2 × 2 × 2 k-mesh. Both MD simulations were performed for 500 fs under the microcanonical ensemble (*NVE*) with the same initial velocities and a time step of 0.5 fs. As shown in Fig. 8(a), the energy fluctuation predicted by the GAP follows closely with that by the AIMD simulation. We further explore the transferability of the GAP by predicting the energies of deformed bulk h-BN. We applied the same strain varying from -5% to +5% to each lattice vector to deform the primitive cell of bulk h-BN, and calculated the potential energy versus volume [$E(V)$] curves using the GAP and DFT. Figure 8(b) shows the $E(V)$ curves predicted by the GAP and DFT agree well with each other. The performance of GAP in reproducing the MD trajectory and $E(V)$ curve demonstrates its high generalizability.

### D. Computational Efficiency of the GAP

Finally, we compare the computational efficiency of our GAP and DFT. We perform the AIMD simulations in VASP and classical MD simulations based on the GAP in LAMMPS using a supercell of bulk h-BN containing 200 atoms. The AIMD and MD simulations are performed in the *NVT* ensemble at 300 K with a time step of 0.5 fs on one Intel Xeon Gold 6254 CPU (18 CPU cores). The computational time per time step of the MD simulation is 0.7s, which is over 2 and 3 orders of magnitude faster than the AIMD simulations using $1 \times 1 \times 1$ (111.2s) and $2 \times 2 \times 2$ (1742.9s) k-meshes, respectively. Although our GAP is remarkably more efficient than DFT calculations, it is still much slower than empirical potentials, which typically are 6-8 orders of magnitude faster than *ab initio* calculations [62]. Further improving the computational efficiency of GAP is thus required to expand its space and time scales for atomistic simulations.

## IV. CONCLUSION

In this work, we report a GAP-type MLIP that can accurately predict the anisotropic thermal properties of bulk h-BN. We determine the appropriate cutoff radius, $n_{max}$, $l_{max}$, and the number of training configurations for constructing a high-quality GAP with high efficiency and accuracy by testing the convergence of the RMSEs of atomic forces and the computational cost. Particularly, we find that including virial stresses in the training data can greatly improve the accuracy of the GAP model for predicting lattice parameters and thermal properties of bulk h-BN.

The developed GAP can well reproduce the overall and spectral thermal conductivities of bulk h-BN with a DFT-level accuracy. Also, the GAP can accurately predict the phonon transport details including the volumetric heat capacity, three-phonon scattering rates, and phonon dispersion. All these results demonstrate that the GAP can well predict the harmonic and anharmonic IFCs, indicating its high precision in reproducing the subtle features of anisotropic PESs of bulk h-BN. Moreover, the GAP exhibits high transferability in predicting the MD trajectory and $E(V)$ curve. Meanwhile, MD simulations based on the GAP can be 2-4 orders of magnitude faster than DFT calculations. Our study demonstrates the high accuracy of the GAP-type MLIP in describing the anisotropic lattice structures and thermal properties of bulk h-BN. Simulations based on MLIPs can be a promising tool for understanding phonon transport and predicting the thermophysical properties of layered materials.

**Acknowledgments**

We acknowledge support from the Excellent Young Scientists Fund (Overseas) of Shandong Province (2022HWYQ-091) and the Initiative Research Fund of Shandong Institute of Advanced Technology (2020107R03). This work used the research computing facilities at Shandong Institute of Advanced Technology.

**REFERENCES**

[1] G. Fugallo, A. Cepellotti, L. Paulatto, M. Lazzeri, N. Marzari, and F. Mauri, Nano Letters **14**, 6109 (2014).

[2] A. Minnich, Physical Review B **91**, 085206 (2015).

[3] C. Chiritescu, D. G. Cahill, N. Nguyen, D. Johnson, A. Bodapati, P. Keblinski, and P. Zschack, Science **315**, 351 (2007).

[4] P. Klemens and D. Pedraza, Carbon **32**, 735 (1994).

[5] W. Jang, Z. Chen, W. Bao, C. N. Lau, and C. Dames, Nano letters **10**, 3909 (2010).

[6] M. Inagaki, Y. Kaburagi, and Y. Hishiyama, Advanced Engineering Materials **16**, 494 (2014).

[7] P. Jiang, X. Qian, R. Yang, and L. Lindsay, Physical Review Materials **2**, 064005 (2018).

[8] M. Raza, A. Westwood, C. Stirling, and R. Ahmad, Composites Science and Technology **120**, 9 (2015).

[9] J. Han, G. Du, W. Gao, and H. Bai, Advanced Functional Materials **29**, 1900412 (2019).

[10] S. K. Nayak, S. Mohanty, and S. K. Nayak, High Performance Polymers **32**, 506 (2020).

[11] Q. Gao, Y. Pan, G. Zheng, C. Liu, C. Shen, and X. Liu, Advanced Composites and Hybrid Materials **4**, 274 (2021).

[12] J. Pei, B. Cai, H. L. Zhuang, and J. F. Li, National Science Review **7**, 1856 (2020).

[13] R. Guo, X. Wang, Y. Kuang, and B. Huang, Physical Review B **92**, 115202 (2015).

[14] Z. Chen, X. Shi, L. Zhao, and J. Zou, Progress in Materials Science **97**, 283 (2018).

[15] Y. Liu *et al.*, Advanced Energy Materials **6**, 1502423 (2016).

[16] B. Yang, L. Hu, W. Ping, R. Roy, and A. K. Gupta, Journal of Energy Resources Technology **144** (2022).

[17] S. E. Kim *et al.*, Nature **597**, 660 (2021).

[18] R. O. Jones, Reviews of Modern Physics **87**, 897 (2015).

[19] G. C. Sosso, J. Chen, S. J. Cox, M. Fitzner, P. Pedevilla, A. Zen, and A. Michaelides, Chemical Reviews **116**, 7078 (2016).

[20] A. K. Rappé, C. J. Casewit, K. Colwell, W. A. Goddard III, and W. M. Skiff, Journal of the American Chemical Society **114**, 10024 (1992).

[21] J. Tersoff, Physical Review B **39**, 5566 (1989).

[22] D. A. Broido, M. Malorny, G. Birner, N. Mingo, and D. Stewart, Applied Physics Letters **91**, 231922 (2007).

[23] L. Lindsay, C. Hua, X. Ruan, and S. Lee, Materials Today Physics **7**, 106 (2018).

[24] W. Li, J. Carrete, N. A. Katcho, and N. Mingo, Computer Physics Communications **185**, 1747 (2014).

[25] T. Feng, L. Lindsay, and X. Ruan, Physical Review B **96**, 161201 (2017).

[26] V. Varshney, S. S. Patnaik, C. Muratore, A. K. Roy, A. A. Voevodin, and B. L. Farmer, Computational Materials Science **48**, 101 (2010).

[27] S. Chen, Q. Wu, C. Mishra, J. Kang, H. Zhang, K. Cho, W. Cai, A. A. Balandin, and R. S. Ruoff, Nature Materials **11**, 203 (2012).

[28] S. G. Volz and G. Chen, Applied Physics Letters **75**, 2056 (1999).

[29] V. L. Deringer, N. Bernstein, A. P. Bartók, M. J. Cliffe, R. N. Kerber, L. E. Marbella, C. P. Grey, S. R. Elliott, and G. Csányi, The Journal of Physical Chemistry Letters **9**, 2879 (2018).

[30] A. J. McGaughey and M. Kaviany, Advances in Heat Transfer **39**, 169 (2006).

[31] A. P. Bartók, M. C. Payne, R. Kondor, and G. Csányi, Physical Review Letters **104**, 136403 (2010).

[32] A. V. Shapeev, Multiscale Modeling & Simulation **14**, 1153 (2016).

[33] A. P. Thompson, L. P. Swiler, C. R. Trott, S. M. Foiles, and G. J. Tucker, Journal of Computational Physics **285**, 316 (2015).

[34] J. Behler and M. Parrinello, Physical Review Letters **98**, 146401 (2007).

[35] V. Botu and R. Ramprasad, Physical Review B **92**, 094306 (2015).

[36] T. D. Huan, R. Batra, J. Chapman, S. Krishnan, L. Chen, and R. Ramprasad, NPJ Computational Materials **3**, 1 (2017).

[37] X. Qian, S. Peng, X. Li, Y. Wei, and R. Yang, Materials Today Physics **10**, 100140 (2019).

[38] H. Babaei, R. Guo, A. Hashemi, and S. Lee, Physical Review Materials **3**, 074603 (2019).

[39] X. Gu and C. Zhao, Computational Materials Science **165**, 74 (2019).

[40] Z. Liu, X. Yang, B. Zhang, and W. Li, ACS Applied Materials & Interfaces **13**, 53409 (2021).

[41] Y. B. Liu, J. Y. Yang, G. M. Xin, L. H. Liu, G. Csányi, and B. Y. Cao, The Journal of Chemical Physics **153**, 144501 (2020).

[42] A. Rodriguez, S. Lam, and M. Hu, ACS Applied Materials & Interfaces **13**, 55367 (2021).

[43] Y. Zhang, C. Shen, T. Long, and H. Zhang, Journal of Physics: Condensed Matter **33**, 105903 (2020).

[44] S. Grenadier, A. Maity, J. Li, J. Lin, and H. Jiang, Applied Physics Letters **115**, 072108 (2019).

[45] A. Maity, S. Grenadier, J. Li, J. Lin, and H. Jiang, Applied Physics Letters **116**, 142102 (2020).

[46] T. T. Tran *et al.*, Physical Review Applied **5**, 034005 (2016).

[47] A. Sajid, M. J. Ford, and J. R. Reimers, Reports on Progress in Physics **83**, 044501 (2020).

[48] C. K. Maity, S. Sahoo, K. Verma, A. K. Behera, and G. C. Nayak, New Journal of Chemistry **44**, 8106 (2020).

[49] Y. Kubota, K. Watanabe, O. Tsuda, and T. Taniguchi, Science **317**, 932 (2007).

[50] S. Ohtani, T. Yano, S. Kondo, Y. Kohno, Y. Tomita, Y. Maeda, and K. Kobayashi, Thin Solid Films **546**, 53 (2013).

[51] Y. Wu *et al.*, ACS Applied Materials & Interfaces **9**, 43163 (2017).

[52] P. Rowe, G. Csányi, D. Alfè, and A. Michaelides, Physical Review B **97**, 054303 (2018).

[53] A. P. Bartók, J. Kermode, N. Bernstein, and G. Csányi, Physical Review X **8**, 041048 (2018).

[54] G. Kresse and J. Furthmüller, Physical Review B **54**, 11169 (1996).

[55] W. Kohn and L. J. Sham, Physical Review **140**, A1133 (1965).

[56] G. Kresse and J. Furthmüller, Computational Materials Science **6**, 15 (1996).

[57] O. Hellman and D. A. Broido, Physical Review B **90**, 134309 (2014).

[58] R. Guo, P. Jiang, T. Tu, S. Lee, B. Sun, H. Peng, and R. Yang, Cell Reports Physical Science **2**, 100624 (2021).

[59] T. Feng, X. Wu, X. Yang, P. Wang, L. Zhang, X. Du, X. Wang, and S. T. Pantelides, Advanced Functional Materials **30**, 1907286 (2020).

[60] A. M. Goryaeva, J. B. Maillet, and M. C. Marinica, Computational Materials Science **166**, 200 (2019).

[61] F. L. Thiemann, P. Rowe, E. A. Müller, and A. Michaelides, The Journal of Physical Chemistry C **124**, 22278 (2020).

[62] C. W. Rosenbrock, K. Gubaev, A. V. Shapeev, L. B. Pártay, N. Bernstein, G. Csányi, and G. L. Hart, npj Computational Materials **7**, 1 (2021).

**TABLE I.** The hyperparameters for the GAP models training.

| Hyperparameters | 2b | 3b | SOAP |
|---|---|---|---|
| $\delta$ (eV) | 10.0 | 3.7 | 0.07 |
| Sparse method | Uniform | Uniform | CUR |
| Sparse points | 50 | 200 | 650 |
| $r_{cut}$ (Å) | 4.5 | 4.5 | 4.5 |
| $\Delta r$ (Å) | - | - | 1 |
| $n_{max}$ | - | - | 8 |
| $l_{max}$ | - | - | 8 |
| $\sigma_v^{energy}$ (eV/atom) | | 0.001 | |
| $\sigma_v^{force}$ (eV/Å) | | 0.0005 | |
| $\sigma_v^{virial}$ (eV/Å$^3$) | | 0.001 | |

**TABLE II.** Lattice constants of bulk h-BN predicted by the GAP-noStress and GAP-Stress, in comparison with the DFT benchmarks.

| Lattice constant | DFT (Å) | GAP-noStress (Å) | Error (%) | GAP-Stress (Å) | Error (%) |
|---|---|---|---|---|---|
| $a$ | 2.49 | 2.48 | 0.14 | 2.49 | 0.04 |
| $c$ | 6.44 | 6.67 | 3.57 | 6.44 | 0.02 |

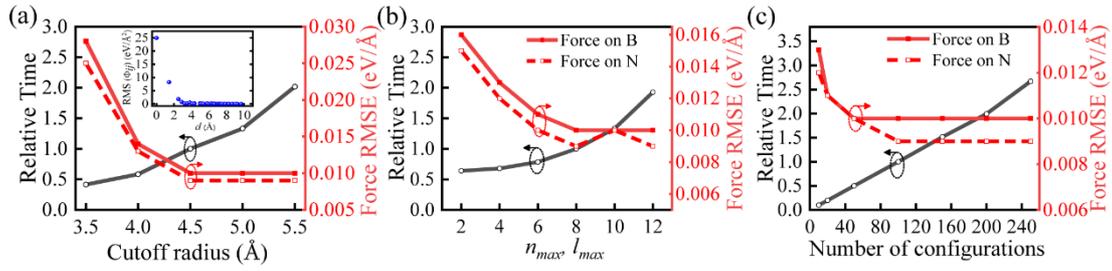

**FIG. 1. Convergence tests of the GAP models without including virial stresses.** RMSEs of atomic forces and relative computational time as a function of (a) the cutoff radius, (b) $n_{max}$ ($=l_{max}$), and (c) the number of the training configurations. Inset: the RMS of the second-order IFC tensors from DFT as a function of the interatomic distance.

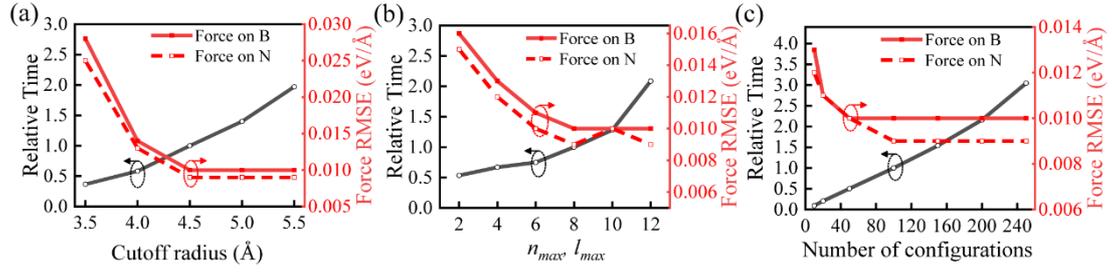

**FIG. 2. Convergence tests of the GAP models including virial stresses.** RMSEs of atomic forces and relative computational time as a function of (a) the cutoff radius, (b) $n_{max}$ ($=l_{max}$), and (c) the number of the training configurations.

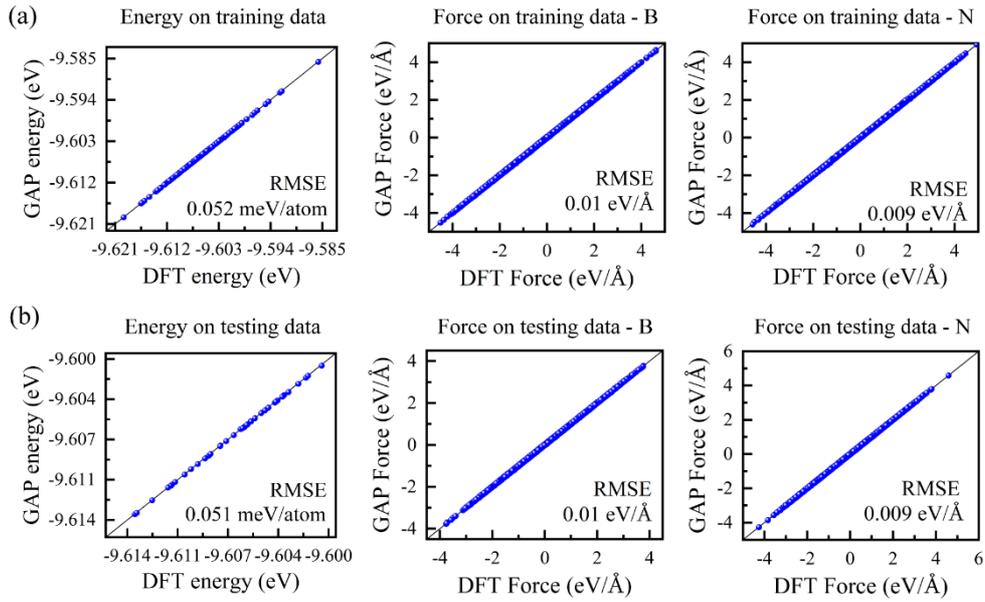

**FIG. 3. Prediction accuracy of the GAP for the total energies and atomic forces on B and N atoms of bulk h-BN.** GAP versus DFT predictions for the (a) training and (b) testing databases.

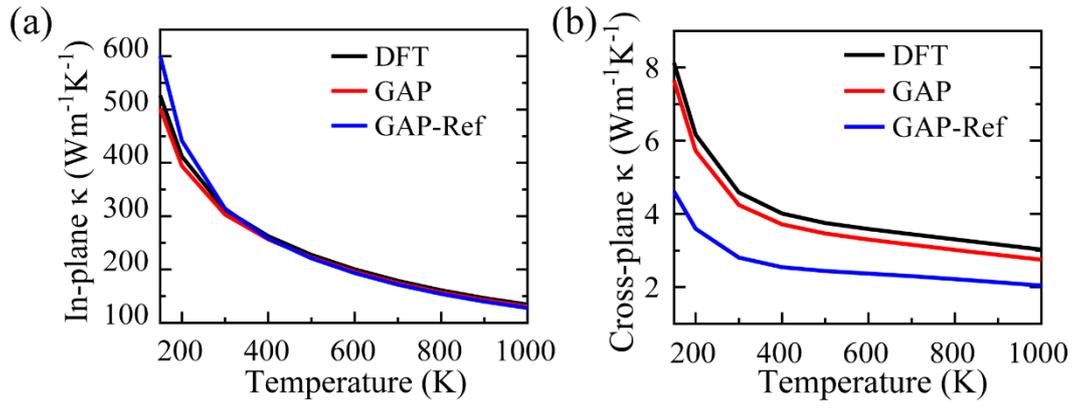

**FIG. 4. Comparison of thermal conductivity of bulk h-BN.** (a) In-plane and (b) cross-plane thermal conductivities of h-BN as a function of temperature predicted by the GAP, GAP-Ref, and DFT.

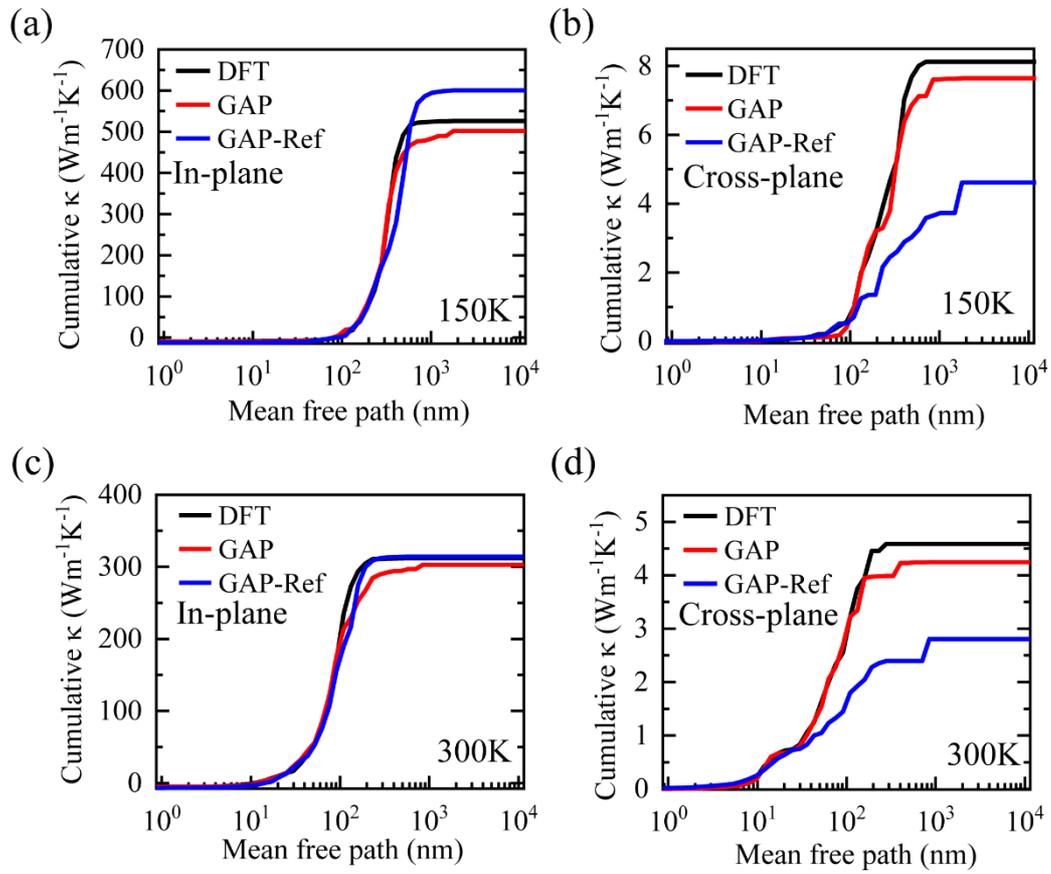

**FIG. 5. Comparison of spectral thermal conductivity for bulk h-BN predicted by the GAP, GAP-Ref, and DFT.** Cumulative thermal conductivities as a function of phonon MFP along (a, c) in-plane and (b, d) cross-plane directions at 150 and 300 K, respectively.

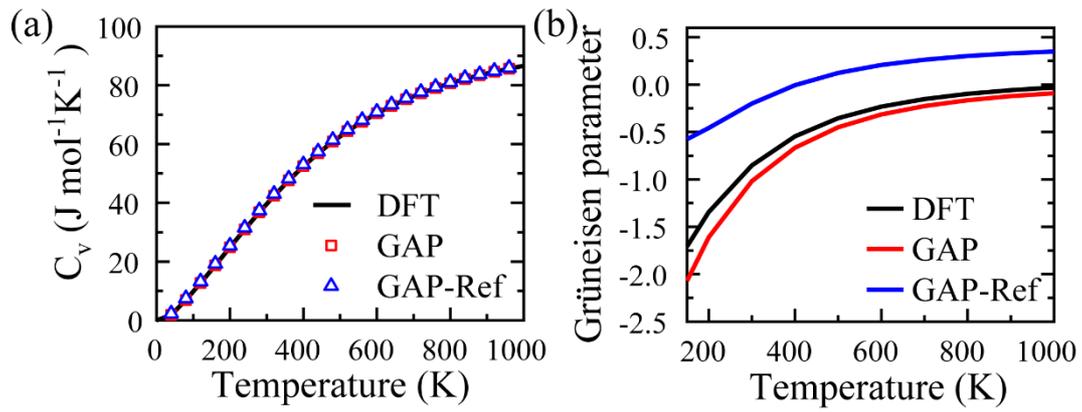

**FIG. 6. Prediction accuracy of thermophysical properties for bulk h-BN.** Temperature-dependent (a) volumetric heat capacity and (b) Grüneisen parameter of bulk h-BN predicted by the GAP, GAP-Ref, and DFT.

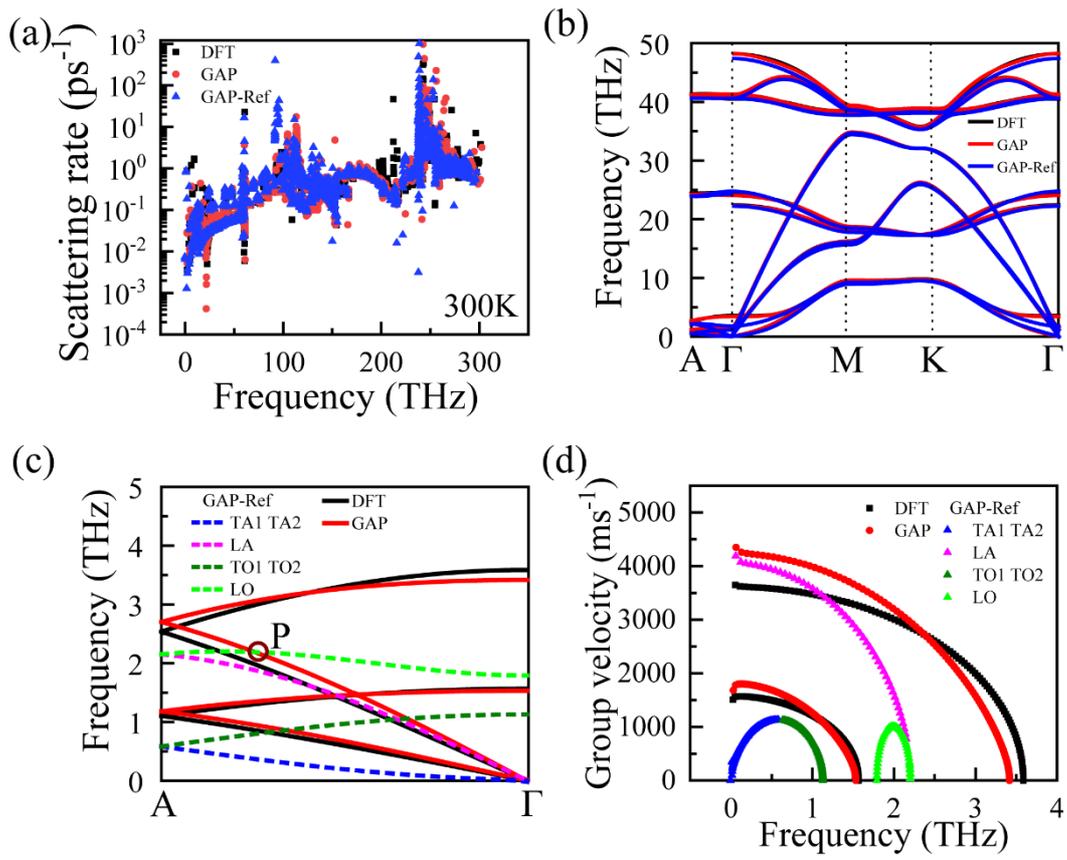

**FIG. 7. Comparison of phonon properties of bulk h-BN predicted by the GAP, GAP-Ref, and DFT.** (a) Three-phonon scattering rates at 300K, and (b) phonon dispersions. (c) Phonon dispersions and (d) group velocities of the six lowest phonon branches along the ΓA direction.

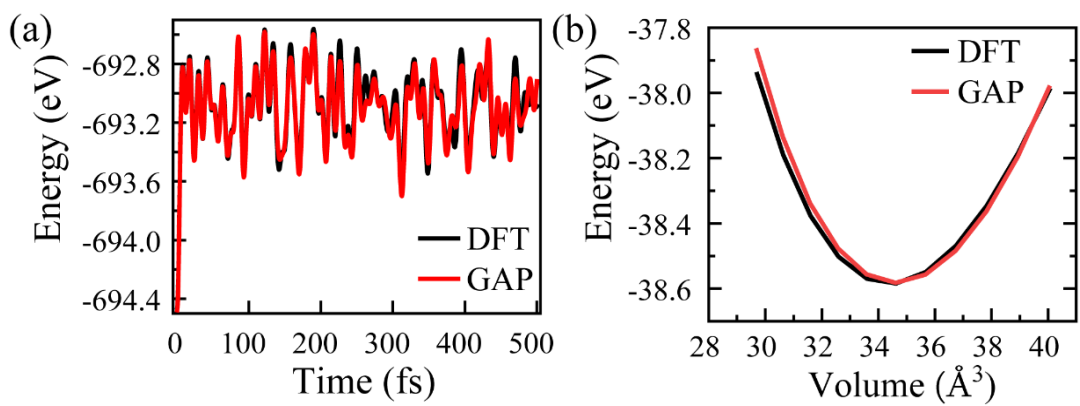

**FIG. 8. Transferability test of the GAP.** (a) The trajectory of potential energy generated by AIMD and MD simulations based on the GAP. (b) Energy-versus-volume $E(V)$ curves calculated by the GAP and DFT.